\documentstyle[graphicx,aps,pre,epsfig]{revtex}
\newcommand {\e}{{\rm e}}

\begin{document}
\title{Replica bounds for diluted non-Poissonian spin systems}

\author{Silvio Franz~\cite{sf},  Michele Leone~\cite{ml}, and
Fabio Lucio Toninelli~\cite{ft}}

\address{ \cite{ml} Institute for Scientific Interchange, viale
Settimio Severo 65, Torino, Italy. \\ \cite{sf} The Abdus Salam
International Center for Theoretical Physics, Condensed Matter Group\\
Strada Costiera 11, P.O. Box 586, I-34100 Trieste, Italy\\ 
\cite{ft} EURANDOM, Technische Universiteit Eindhoven, P.O. Box
513, 5600 MB Eindhoven, The Netherlands\\}

\date{\today} 
\maketitle

\begin{abstract}
In this paper we extend replica bounds and free energy subadditivity 
arguments to diluted spin-glass models on graphs with arbitrary,
non-Poissonian degree distribution. The new difficulties specific of
this case are overcome introducing an interpolation procedure that 
stresses the relation between interpolation methods and the cavity
method. As a byproduct we obtain self-averaging identities that
generalize the Ghirlanda-Guerra ones to the multi-overlap case. 
\end{abstract}

\section{Introduction}

The replica and the cavity methods play a fundamental role in the
analysis of mean field disordered systems, where they suggest a low
temperature glassy phase with a great extent of universality \cite{PMV}.

With the technique of ``replica symmetry breaking'' (RSB)
physicists have been since a long time able to describe in great
detail the statistical properties of the configurational space of
``fully connected'' disordered models, where each degree of freedom
interacts with an extensive number of neighbours.

The last times have seen important progresses in extending the
analysis to ``diluted models'', where each degree of freedom interacts
with a finite number of randomly chosen neighbours. Although the
equations describing RSB in these models were well known already in
the past, it is only with the recent introduction of a population
dynamics algorithm \cite{MP-pop,MP} that it has been possible to tackle the
subtleties of the glassy low temperature phases.  Thanks to this
algorithm and its specification to zero temperature it has been
possible to analyze at the level of ``one step replica symmetry
breaking'' (1RSB) a great variety of problems relevant for physics of
glassy systems or computer science or both. These range from the
Viana-Bray model for spin glasses \cite{MP}, where the 1RSB is
expected to give just an approximate result, to the diluted $p$-spin
model - or XOR-SAT problem \cite{FEDERIME} -, the
Biroli-M\'ezard lattice glass model \cite{LATTICEGAS}, models for
efficient error correcting codes \cite{codici}, the random K-SAT
problem and graph coloring \cite{MPZ,PRE,coloring},
where at least at zero temperature  1RSB is thought to be exact.

The population dynamic algorithm, originally thought for computing
averages over random graph disorder, has been suitably generalized to
deal with single samples, and in conjunction with decimation
algorithms has been proved to be effective to solve random K-SAT or
graph coloring in essentially linear time up to very close to the
SAT/UNSAT transition  threshold \cite{PRE}.

Despite these successes, together with a large wealth of
important rigorous results on both mean field and short range models
\cite{matematici,tala-p-spin,ShortCourse}, 
the theoretical foundations of replica and
cavity methods remain unsatisfactory. On one hand replica methods rely
on a non-unique analytic continuation of the integer moments of the
partition function which is difficult to control mathematically; on
the other, the cavity method is based on physically sound, but
unproven assumptions on the nature of the low temperature pure
states. 

A great deal of mathematical work has been devoted to the analysis of
disordered models with rigorous -i.e., conventional probabilistic-
methods, which in many cases allow to control the RS high
temperature case, and in the remarkable analysis of Talagrand 
\cite{tala-p-spin} for the
fully connected $p$-spin for large enough $p$, even the low temperature
one. A different approach has been put forward by Guerra 
\cite{guerra}, who was able to prove that in a large class of long
range models the free energy can be written as the sum of the replica
expression plus a reminder which, by simple inspection, is proven to
be positive. This was the first proof that the replica/cavity Ansatz
is in some cases a variational Ansatz, therefore justifying the
replica rule that the free energy should be maximized with respect to
the replica parameters.\footnote{Cases are know in replica theory
where the right free energy extremum is a saddle point, but this does
not happen in the mentioned cases.}  The technique of Guerra is based
on the idea of interpolating the original model with a pure paramagnet
with suitably chosen local external fields. In the original
formulation the interpolation was performed tuning the relative
strengths of the couplings and the external fields, and relayed heavily
on the Gaussian and long range nature of the interactions among spins.
This technique has been employed in several papers dealing with 
long and short range models \cite{contucci0}.

In a previous paper \cite{primolavoro}, hereafter referred to as I, two
of us have shown how to generalize the technique of the interpolating
model to diluted models on graphs with Poissonian degree
distribution. In that case, the interpolation is done progressively
removing addends in the Hamiltonian, and compensating in average by
the introduction of appropriate external fields on random sites of the
lattice. The method was then interpreted as a variational version of
the cavity method. The result was that the free energy could be
written as a sum of a replica/cavity like contribution plus a
remainder.  For the even $p$ diluted $p$-spin model and the even $K$
$K$-SAT problems, simple considerations not relying on the actual
solutions to the replica/cavity equations, allowed to conclude that
the remainder was positive. This means that the replicas give
free energy lower bounds to these models. Specified to zero
temperature, this proves that replicas give an upper bound for the
satisfiability threshold, that is the value of the degree of
connectivity below which all the terms of the Hamiltonian can be
minimized at the same time. The cases of odd $p$ or $K$ led to an
expression for the reminder whose sign cannot be obviously decided,
thus leaving as an open question if the recently obtained 1RSB
solution for the random $3$-SAT problems or the odd $p$-spin case
actually give free energy lower bounds and SAT-threshold upper bounds.

As we said, the results were confined to models on Erd\"os-R\'enyi
(hyper)-graphs which have Poissonian degree distribution. However, the
replica and the cavity methods have been successfully applied for spin
models on more general graphs, where the degree distribution is
basically arbitrary.

Motivated by that consideration, in this paper we generalize the
analysis to random graphs with arbitrary connectivity
distribution. Here, differently from the Poissonian case, the
compensation in average is not possible and the erasure of a term in
the Hamiltonian requires detailed compensation through the
introduction of proper fields on the sites belonging to the erased
clause. The new difficulties arising in this case can be overcome
employing the property of self-averaging of appropriate observables,
leading to identities on multi-overlap distributions that generalize
the Ghirlanda-Guerra identities \cite{GG} for the usual overlaps.
This new feature is common both to the RS and the RSB case.  For
exposition simplicity we will limit ourselves in this paper to the
discussion of the RS case.

The paper is organized as follows: in section 2 we define the model
and introduce the notations, in section 3 we introduce the
interpolating models, which we use in sections 4 and 5 to find
representations of the free energy as main terms plus remainders,
suitable respectively to prove replica bounds and subadditivity. In
section 6 we use the self-averaging to estimate some terms in the
remainders and generalize the Ghirlanda-Guerra identities. Section 7 
is devoted to showing the positivity of the remainders, and finally
in section 8 we discuss some conclusions.

\section{Definition of the models and summary of mathematical notations}

\subsection{The Models}

In this paper we consider diluted spin models on random graphs with arbitrary 
degree distribution, consisting in  a
collection of $N$ Ising $\pm 1$ spins ${\bf S}=\{S_1,...,S_N\}$,
interacting through Hamiltonians of the kind
\begin{equation}
{\cal H}^{(M)}({\bf S},{\bf J})= \sum_{\mu=1}^M
H_{J^{(\mu)}}(S_{i_1^\mu},...,S_{i_p^\mu})
\label{general}
\end{equation}
where the indexes $i_l^\mu$ are i.i.d.  quenched random variables
chosen in the following way: one first extracts a set of site degrees
$k_i$, representing the number of clauses  where $S_i$ appears,
($i=1,...,N$), as i.i.d. from a distribution $p_k$. Then the
configuration of indexes $i_l^\mu$ are chosen uniformly among all the
possible ways respecting the prescribed degrees. In other words, the
joint probability of all the indexes $\{ i_l^\mu\}$ will be
proportional to
\begin{equation}
\prod_{i=1}^N \delta (\sum_{\mu=1}^M \sum_{l=1}^p
\delta_{i_l^\mu,i}-k_i)
\label{indici}
\end{equation}
where both $\delta$'s appearing in (\ref{indici}) denote the Kronecker
symbol. We will concentrate on regular degree
distributions where all the moments $\langle k^l\rangle =\sum_k p_k
k^l$ are finite. The number of clauses $M$ is therefore  itself in principle 
a random variable given by
$M=\frac{1}{p}\sum_{i=1}^N k_i$, its
 average  $\langle M\rangle =\alpha N$ will be proportional to $N$, with
$\alpha=\langle k\rangle/p$ and  it will have small $O(\sqrt{N})$
fluctuations around its average.  For this reason, we will often 
treat $M$ as a constant, making only a $O(1/N)$ error in the free energy.
As it will become soon clear, the
subscript $J^{(\mu)}$ in the clauses indicates dependence on a single
or on a set of quenched random variables.

We will treat in a unitary way the case 
of the the $p$-spin model\cite{ri-we-ze} or random XOR-SAT problem of
computer science, and the random K-SAT model. 
In the $p$-spin model the  clauses have  the form
\begin{equation}
H_{J^{(\mu)}}(S_{i_1^\mu},...,S_{i_p^\mu})=J^\mu S_{i_1^\mu}\cdot
...\cdot S_{i_p^\mu} \;
\label{spin}
\end{equation}
where $J^\mu$ will be taken as i.i.d. random
variable with regular symmetric distribution $\mu(J)=\mu(-J)$. 
Particular attention will be given to Viana-Bray model corresponding
to $p=2$. 
In random K-SAT model the clauses have the form \cite{ksat}
\begin{equation} 
H_{J^{(\mu)}}(S_{i_1^\mu},...,S_{i_p^\mu})= \prod_{l=1}^p
\frac{1+J^\mu_{i_l^\mu}S_{i_l^\mu}}{2} \; ,
\label{SAT}
\end{equation} where the quenched variables 
$J^\mu_{i_l^\mu}=\pm 1$ are i.i.d. with symmetric
probability.\footnote{While the assumption of a symmetric distribution
$\mu(J)=\mu(-J)$ will play an important role
in establishing the free energy bounds, the precise form of the
distribution will not be essential.} The number $p$ of spins appearing
in a clause is usually called $K$ in the K-SAT problem, but for uniformity
of notation we will deviate from this convention.

In I the indexes $i_l^\mu$ of the
spins appearing in the clauses were chosen with uniform probability,
giving rise to random graphs with Poissonian degree statistics, and
the treatment was based on the peculiar property of the Poisson
distribution. Here we show that thanks to  self-averaging of certain
quantities, the validity of the results does not depend on the
specific form of the graph degree distribution. 

\subsection{Notations}

Let us establish some notations. We will need several kinds of
averages:
\begin{itemize}
\item The Boltzmann-Gibbs average for fixed  quenched disorder: given
an observable $A({\bf S})$
\begin{equation}
\omega(A)=\frac{\sum_{{\bf S}}A({\bf S}) \exp(-\beta {\cal H}({\bf
S},{\bf J}))}{Z}
\end{equation}
where $Z={\sum_{{\bf S}}\exp(-\beta {\cal H}({\bf S},{\bf J}))}$ and
$\beta$ is the inverse temperature.

Obviously, $\omega(A)$, as well as $Z$ will be functions of the
quenched variables, the size of the system and the temperature. This
dependence will be  made explicit only when needed.

%Finally, the disorder-dependent free energy $F_N(\beta,J)$ and its quenched
%average
%$F_N(\beta)$ are defined as
%\begin{eqnarray}
%  &&F_N(\beta,J)=-\frac1{N\beta} \ln Z(\beta,J)\\\nonumber
%&&F_N(\beta)= E F_N(\beta,J).
%\end{eqnarray}

\item
The disorder average: given an observable quantity $B$ dependent on
the quenched variables appearing in the Hamiltonian,  we will denote
as $E(B)$ its average. This will include the average with respect to
the $J$ variables and the choice  of the random indexes in the clauses
as well as with respect to other  quenched variables to be introduced
later.
\item
We will need in several occasions the ``replica measure''
\begin{equation}
\Omega (A_1,...,A_n)=\omega(A_1)...\omega(A_n).
\label{repmeas}
\end{equation}
\item

Moreover, throughout the work we will consider modified versions of
the original Hamiltonians that will depend on a discrete
dilution parameter $t$. Both Boltzmann and disorder  averages will
depend accordingly on $t$.
%, and will be denoted with an apex $(t)$. 
The
original averages will correspond to $t = M$.
\item
Another notation we will have the occasion to use is the one  for the
overlaps among $l$ spin configurations $\{ S_i^{a_1},...,S_i^{a_l}
\}$, out of a population of $n$   $\{ S_i^{1},...,S_i^{n} \}$:
\begin{equation}
q^{(a_1,...,a_l)}=\frac 1 N \sum_{i=1}^N S_i^{a_1} \cdot ... \cdot
S_i^{a_l} \;\;\;\; (1\le a_r \le n \;\;\; \forall r) ,
\end{equation} 
and in particular
\begin{equation}
q^{(n)}=q^{(1,...,n)}=\frac 1 N \sum_{i=1}^N S_i^{1} \cdot ... \cdot
S_i^{n}.
\end{equation} 
\item Finally, we will denote as $T_i$ the set of clause indexes where  
the $i'th$ spin appears and $V_\mu$ the set of spin indexes belonging to the
clause $\mu$. 
\end{itemize}
In the following we will need to consider averages where some of the
variables are excluded, e.g., the averages when a variable  $u_i^{k_i}$
is erased.  These averages will be denoted with a subscript
$-u_i^{k_i}$, e.g., if an $\omega$ average is concerned the notation
will be $\omega_{-u_i^{k_i}}(\cdot)$.  Other notations will be defined
later in the text whenever needed.

Our interest will be confined to bounds to the free energy density
$F_N = -\frac{1}{\beta N} E \log Z$ and the ground state energy
density $U_{GS} =  \lim_{N \to \infty} 1/N E \left[ min \, (U_N)
\right] $ valid in the thermodynamic limit, so that $O(1/N)$ will be
often implicitly neglected in our  calculations.

\section{Introducing the interpolating models}

We will use the technique of interpolating models for the purposes
of showing the existence of the infinite volume free energy, and
of proving replica bounds. For exposition reasons, we will explain 
the method first for the replica bounds and later for the existence 
of the free energy. 

\subsection{The replica/cavity bounds: clause deletion versus fields 
compensation}

\begin{figure}
\centering \epsfxsize=\textwidth \epsffile{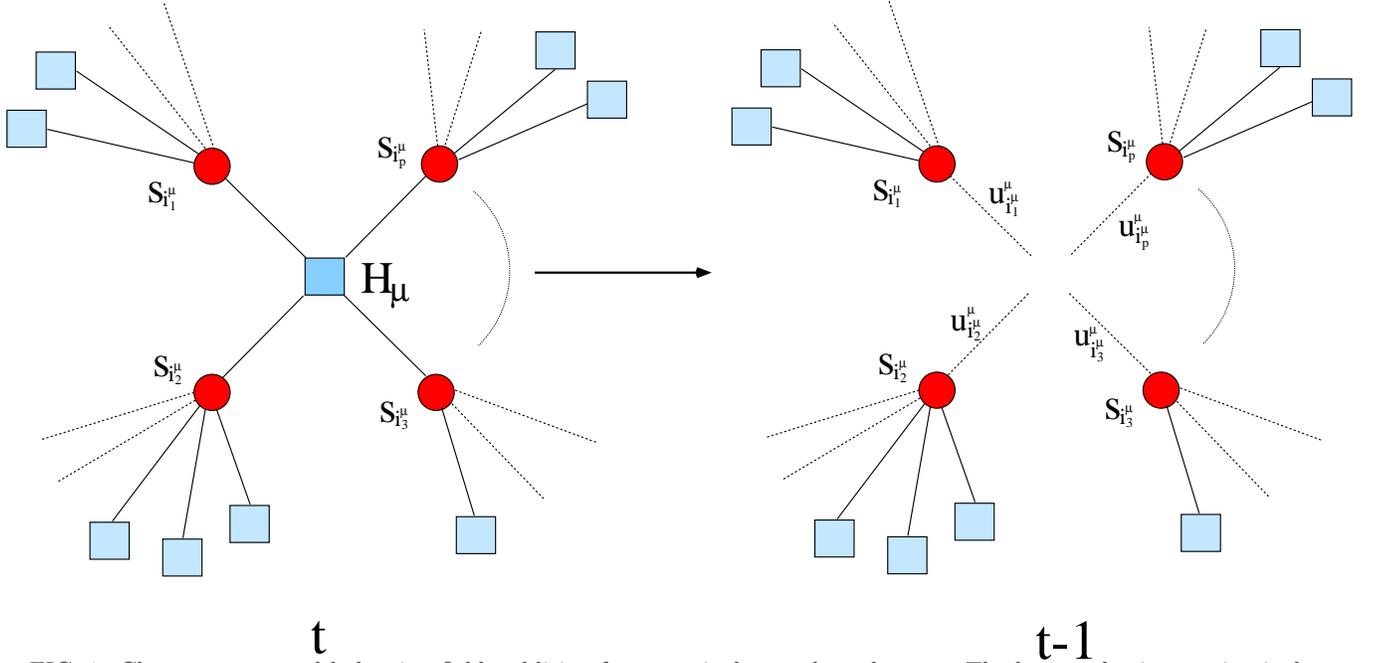}
\caption{Clause erasure and balancing fields addition for a particular
random clause $\mu$. The hyper-edge interaction is drawn in the
{factor-graph notation} with square nodes representing clauses and
circles representing spins. Dotted lines represent the added balancing fields. 
Only clause $\mu$ and corresponding variables and fields are explicitly named.}
\label{figura2}
\end{figure}

In order to prove the free energy bounds, we will use an iterative
discrete graph pruning procedure where at each time step $t$
decreasing from $M$ to $0$, we erase the clause labelled by $t$ and
compensate this reduction by the introduction of some auxiliary fields
$u_{i_l^t}^t$ on the sites $i_l^t$ ($l=1,...,p$) belonging to the
clause $t$ (see Fig.(\ref{figura2})). At each time step $t$ we have a
different model with Hamiltonian
\begin{equation}
{\cal H}^{(t)}[{\bf S}]=\sum_{\mu=1}^t 
H_{J^{(\mu)}}(S_{i_1^\mu},...,S_{i_p^\mu})-\sum_{\mu=t+1}^M \sum_{l=1}^p
u_{i_l^\mu}^\mu S_{i_l^\mu}
\label{compoundH}
\end{equation}
which interpolates between a simple paramagnet made of non interacting
spins at time $0$ and the original model at time $M$.  The clause
removal procedure is very similar to the analogous operation of spin
and clause addition usually performed in the cavity method, the
difference being that here the deletion is explicitly compensated by the
introduction of external fields. A key point in obtaining the
free energy replica/cavity bounds is to assume that the {\it external}
fields are random variables obeying the statistics of the {\it cavity}
fields in the cavity approach. 

In order to explain this point, and to motivate the introduction
of the definitions below, let us remind some formulae of the
cavity approach. In that context,  one singles
out the contribution of the clauses and the sites to the free energy
and defines cavity fields $h_i^{\mu}$ and $u^\mu_{i}$ respectively
as the local field acting on the spin $i=i_1^\mu$ in absence of the clause
$\mu$ and the local field acting on $i$ due to the presence of the
clause $\mu$ only. If we define
$Z[S_{i}]$ as
the partition function of
a given sample with $N$ spins where all but the spin $S_i$ are
integrated,  and $F_{N,-i}$ as the free energy of the
corresponding systems where the spin $S_i$ and all the clauses which 
belong to it are removed, 
in the cavity approach one assumes that 
\begin{eqnarray}
{ Z}[S_i]& \approx & e^{-\beta { F}_{N,-i}} 
\prod_{\mu\in T_i}
\sum_{S_{i_2^\mu},...,S_{i_p^\mu}} e^{-\beta
H_{J^{(\mu)}}(S_{i},S_{i_2^\mu},...,S_{i_p^\mu})+\beta \sum_{l=2}^{p}
h_{i_l^\mu}^{\mu}S_{i_l^\mu} } \nonumber\\  & = & e^{-\beta
{ F}_{N,-i}} \prod_{\mu\in T_i}  B_{\mu}^{(i)}  
e^{\beta u^\mu_{i} S_{i}}.
\label{c1}
\end{eqnarray}
Notice that in general, if (\ref{c1}) had to represent the exact
integration of $N-1$ spins in 
in a finite $N$ systems,
effective $l$-spin interactions ($l=2,...,p$) should also be present among
the spins explicitly appearing in the r.h.s. of (\ref{c1}). Given the
peculiar topology of random graphs, where loops of interacting
variables have generically length $O(\log(N))$, these coupling can be
expected to be small and are neglected in the approach. 
This is the reason why we wrote $\approx$ instead of $=$.  The
constant $B_{\mu}^{(i)}=e^{-\beta \Delta F_{\mu}^{(i)}}$ is 
interpreted as a suitable shift in the free energy due to the
contribution of the clause $\mu$ for fixed value of the spin $i$.  We
notice that denoting $J^{\mu}$ as $J$, and renaming the fields  in
(\ref{c1}) into $h_1,...,h_{p-1}$, Eq. (\ref{c1}) defines functions
\begin{equation}
u_J(h_1,...,h_{p-1}) \; \;  {\rm and} \; \; B_J(h_1,...,h_{p-1}) \; 
\label{ub}
\end{equation} 
that one can compute explicitly using the form of the clauses of the 
different models. For instance, in the case of the Viana-Bray model, where 
$p=2$, one has:
\begin{eqnarray}
u_J(h)&=&\frac{1}{\beta} \tanh^{-1}[\tanh(\beta J)\tanh(\beta h) ] \\
B_J(h)&=& \frac{2 \cosh(\beta J)\cosh(\beta h)}{\cosh(\beta u_J(h))}.
\label{ub1}
\end{eqnarray} 

Always in the cavity approach one closes the set of 
equation imposing the self-consistent equations
\begin{equation}
h_{i}^{(\mu)}=\sum_{\nu \in \{ T_i-\mu\}} u_\nu^{(i)}. 
\label{c2}
\end{equation}

Following the intuition based on Eq. (\ref{c1}) we will define the
external fields $u_{i_l^\mu}^\mu$ in the
interpolating model as verifying the relation 
\begin{equation}
u_{i_l^\mu}^\mu=u_{J^{(\mu)}}(g_{i_1^\mu}^\mu,...,g_{i_{l-1}^\mu}^\mu,
g_{i_{l+1}^\mu}^\mu,,...,g_{i_{p}^\mu}^\mu)
\end{equation}
where the $p-1$ arguments $g$ will be independent variables with
suitable statistics. We do not write at this point a relation
analogous to the self-consistency relation (\ref{c2}).  This equation
will be obeyed in average when the statistics of the fields $g$ is
chosen in such a way to optimize the replica bounds.  The key point of
the procedure, consists in the choice of the distribution of the
primary fields $g_{i_l^\mu}^\mu$.  Different free energy bounds can be
obtained assuming the type of statistics implied by the different
replica solutions.
In this paper we concentrate on the replica symmetric bound, that will be
obtained considering the primary fields $g_{i_l^\mu}^\mu$ as
i.i.d. variables drawn from a distribution $G(g)$. 
Correspondingly the $u$'s will be distributed according to $Q(u)$
verifying 
\begin{eqnarray}
Q(u)& = & \int dg_1 \;G(g_1)... dg_{p-1} \;G(g_{p-1})
\langle \delta (u-u_J(g_1,...,g_{p-1}))\rangle_J. 
\label{qdiu}
\end{eqnarray} 
Writing the free energy of the interpolating model as 
\begin{equation}
F_N(t)=-\frac{1}{\beta N}E \log Z(t), 
\end{equation} 
we observe that the average
free energy $F_N=F_N(M)$ of the original model can be written as 
\begin{equation}
F_N  =  \sum_{t=1}^M \Delta_N(t) + F(0)
\label{Ftot}
\end{equation}
where we defined the discrete time derivative  of the free energy as 
\begin{equation}
\Delta_N (t) = F_N(t) - F_N(t-1). 
\label{deltaf}
\end{equation}
The bound will consist in comparing $\Delta_N(t)$ with the corresponding
replica symmetric expression. 

The extension to 1RSB bounds, although technically more involved, does
not present additional conceptual difficulties. In I, it has been
shown how to provide 1RSB bounds in the Poissonian case. One needs to
assume that the fields $g_{i_l^\mu}^\mu$ are still independent
variables, but subject to site and clause distributions
$G_{i_l^\mu}^\mu$ which are themselves random and independent from one
another and subject to the common functional distribution
${\cal{G}}({G})$. The appropriate definition of interpolating
free energy will depend on a parameter $m\in[0,1]$:
\begin{equation}
F_N(t)=-\frac{1}{m \beta N} E_1
\log E_2 Z(t)^m.  
\end{equation} 
We denoted by $E_2$ the average over the fields
$g_{i_l^\mu}^\mu$ for fixed
distributions  $G_{i_l^\mu}^\mu$ and by $E_1$ the average over the
distributions  $G_{i_l^\mu}^\mu$ as well as over all the other
quenched variables. 
For $t=M$ the $E_2$ average is immaterial  and  
formula (\ref{Ftot}) holds. 
For general degree distribution one 
could follow the same procedure, but in order to
keep this paper within a reasonable length we will not present here
this case. 

\subsection{The thermodynamic limit: a cut and paste procedure.}

\begin{figure}
\centering \epsfxsize=1.\textwidth \epsffile{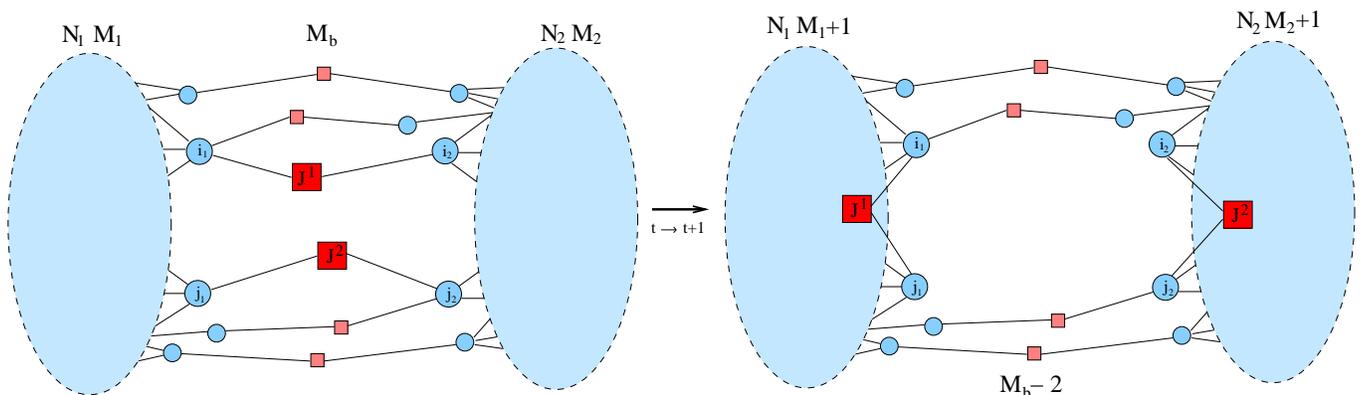}
\caption{Cut and paste procedure in the Viana-Bray model. The groups of constraints 
not belonging to $M_b$ and of graph sites not participating to 
the constraints in $M_b$ are generically represented as dashed lines sets.}
\label{cutandpaste}
\end{figure}

The aim of this section is to interpolate between the original system
of $N$ interacting spins and two separate models, respectively with
$N_1$ and $N_2$ spins ($N_1+N_2=N$) and to show that the free energy
is subadditive \cite{guerraTD}.  For notational simplicity we
consider here explicitly just the case of the Viana-Bray model, with
Hamiltonian $H=-\sum_{\mu=1}^M J^\mu S_{i_1^\mu}S_{i_2^\mu}$. Inspired
by the construction of the previous section we start from the model
with $N$ spins and consider the sets of the first $N_1$ spins and the
one of the remaining $N_2$ spins. Each of the $M$ clauses, will either
belong to the first sub-system, if $i_1^\mu,i_2^\mu\in \{1,...,N_1\}$,
or to the second if $i_1^\mu,i_2^\mu\in \{N_1+1,...,N\}$ or they will
be ``bridge'' clauses if one of the indexes is less or equal than
$N_1$ and the other is greater than $N_1$. Let us denote as
$M_1(0),M_2(0)$ and $M_b(0)$ the number of clauses of the different
types, respectively.  We define our interpolating model via an
iterative ``cut and paste'' procedure where at each time step we
select at random two bridge clauses, we cut them, and we reconnect the
spins belonging the first sub-system between themselves with a new
random coupling, and similarly for the two spins belonging to the
second sub-system (see Fig.(\ref{cutandpaste})). 
In such a way at each time step
$t=1,...,M_b(0)/2$ (we suppose for simplicity to choose the ordering
of the spins in such a way that $M_b(0)$ is even) each spins conserves
its original connectivity, and the number of clauses of the different
kinds are modified as $M_1(t)=M_1(t-1)+1$, $M_2(t)=M_2(t-1)+1$ and
$M_b(t)=M_b(t-1)-2$. At the end of the procedure we have two separate
models with, respectively, $N_1$ spins and $M_1(0)+M_b(0)/2$ clauses and
$N_2$ spins and $M_2(0)+M_b(0)/2$ clauses. Notice that $M_1, M_2$ and
$M_b$ are random variables with average proportional to $N$ and small
$O(\sqrt N)$ fluctuations.  As in the case of $M$, we will neglect
these harmless fluctuations in our analysis.  It is easy to realize
that the graphs associated to the resulting non-interacting models are
chosen with uniform probability among the ones having the correct
connectivities for each spin.  This can be explicitly checked by an
elementary induction calculation. Each step of the cut and paste
procedure transforms the uniform distribution on the graphs with
prescribed connectivities $k_i$ of the sites and number of clauses of
the three types $M_1(t)$, $M_2(t)$ and $M_b(t)$, into the uniform
distribution with the same connectivities but with clause numbers
$M_1(t)+1$, $M_2(t)+1$ and $M_b(t)-2$.

Therefore, in analogy with (\ref{Ftot})  we have 
\begin{equation}
 F_N= \frac {N_1}N F_{N_1}+\frac{N_2}N F_{N_2}- \sum_{t=1}^{M_b(0)/2}\Delta'_N(t),
\label{inanalogy}
\end{equation}
where $\Delta'_N(t)$ is the discrete time derivative along the cut and
paste procedure. 
As will be shown in section 5, $\Delta'_N(t)$ is non-negative for the
models
we are considering, as long as $p$ is even, thus implying
subadditivity of the free energy, and therefore the existence of its
infinite volume limit.

\section{The RS bound}

Let us start with a few observations independent of the chosen
statistics of the primary fields, and therefore valid both for the RS
and for the RSB estimate. 

In order to compare the free-energies $F(t)$ and $F(t-1)$, let us
consider $Z(t-1)$ and isolating the terms containing the fields
$u_{i_l^t}^t$ observe that this can be written as
\begin{equation}
Z(t-1) = Z_{-u^t}(t-1)\omega_{-u^t}^{(t-1)}(\e^{\beta \sum_{l=1}^p
  u_{i_l^t}^t S_{i_l^t}})
\label{zt1}
\end{equation}
where $Z_{-u^t}(t-1)$ and $\omega_{-u^t}^{(t-1)}(\cdot)$ are
respectively the partition function and the Boltzmann averages
corresponding to the Hamiltonian at time $t-1$ in absence of the
fields $u_{i_l^t}^t$. In the same way, isolating the $t$-th clause term
in $Z(t)$ and noticing that $Z_{-J^{(t)}}(t)=Z_{-u^t}(t-1)$ and
$\omega_{-J^{(t)}}^{(t)}(\cdot)=\omega_{-u^t}^{(t-1)}(\cdot)$ we can write
\begin{equation}
Z(t) = Z_{-u^t}(t-1)\omega_{-u^t}^{(t-1)}(\e^{-\beta
H_{J^{(t)}}(S_{i^t_1},...,S_{i^t_p})}).
\label{zt2}
\end{equation}
Using these relation in the definition of the RS free energy time
derivative $\Delta_N(t)$ we find: 
\begin{equation}
N \Delta_N(t)=-T \left[  
E\left(\ln \omega_{-u^t}^{(t-1)}(\e^{-\beta
H_{J^{(t)}}(S_{i^t_1},...,S_{i^t_p})})
\right)
-
E\left(\ln \omega_{-u^t}^{(t-1)}(\e^{\beta \sum_{l=1}^p
  u_{i_l^t}^t S_{i_l^t}})
\right)
\right]
\label{del}
\end{equation}

Now we compare this term with the one 
obtained in the cavity approach (supposing that the statistics of the
cavity fields $h$ coincide with the one of our external primary fields
$g$):  
\begin{equation}
\label{appr1}
\omega_{-u^t}^{(t-1)}(\e^{-\beta
H_{J^{(t)}}(S_{i^t_1},...,S_{i^t_p})})\approx\frac{1}{\prod_{l=1}^p
  2\cosh(\beta h_{i_l^t}^t)}
\sum_{S_1,...,S_p}\e^{-\beta
H_{J^{(t)}}(S_{i^t_1},...,S_{i^t_p})+\beta \sum_{l=1}^p h_{i_l^t}^t S_l}
\end{equation}
and in the same way
\begin{equation}
\label{appr2}
\omega_{-u^t}^{(t-1)}(\e^{\beta \sum_{l=1}^p
  u_{i_l^t}^t S_{i_l^t}}
)\approx \frac{1}{\prod_{l=1}^p
  2\cosh(\beta h_{i_l^t}^t)}
\sum_{S_1,...,S_p}
\e^{\beta \sum_{l=1}^p
  u_{i_l^t}^t S_{l}+\beta \sum_{l=1}^p h_{i_l^t}^t S_l}. 
\end{equation}
This leads to  the expression: 
\begin{eqnarray}
N { \Delta}_N(t)& \approx &-T \left[  
E
\log\left( 2^{-p}
\sum_{S_1,...,S_p}\e^{-\beta
H_{J^{(t)}}(S_{1},...,S_{_p})}
\prod_{l=1}^p (1+S_l \tanh(\beta h_{i_l^t}^t))
\right)\right.\nonumber\\
& &
\left.
-
E\sum_{l=1}^p\left(
\log(\cosh(\beta u_{i_l^t}^t))
+
\log(1+\tanh(\beta h_{i_l^t}^t)\tanh(\beta u_{i_l^t}^t))
\right)
\right].
\label{delvero}
\end{eqnarray}
The replica symmetric approximation to this expression consists in
assuming that the statistics of the cavity fields $h_{i_l^t}^t$
coincide with the one of the external fields $g_{i_l^t}^t$, and we
call $\Delta^*(t)$ the expression corresponding to (\ref{delvero}), 
once these substitutions have been made. 

Now, we go back to our approach where, it is important to emphasize, 
we are not assuming the validity of approximations like (\ref{appr1}, 
\ref{appr2}).
In order to get a control of
the free energy we add and subtract $\Delta^*(t)$ from the expression
(\ref{del}) of $\Delta_N(t)$. It is also useful to add and subtract the
term
\begin{equation}
T\sum_{l=1}^p 
E\log(1+\omega_{-u^t_{i_l^t}}^{(t-1)} (S_{i_l^t})\tanh(\beta u_{i_l^t}^t))
\end{equation}
Rearranging terms in $\Delta_N$ and taking into account that
\begin{equation}
F(0) = -\frac{1}{\beta N} E\sum_{i=1}^N\log ( 2 \cosh (\beta
\sum_{\mu\in T_i}u_i^\mu ))
\end{equation}
we rewrite the free energy as: 
\begin{equation}
F_N = F_{var}[G] + \frac{1}{N}\sum_{t=1}^M R[G,t] + \frac{1}{N}
\sum_{t=1}^M {\tilde
R}[G,t] + O(1/N).
\label{FveraRS}
\end{equation}
Here, we have isolated the ``variational term'' $F_{var}[G]$
\begin{eqnarray}
F_{var}[G]& = &
 -\frac{1}{\beta N} E\sum_{i=1}^N\log ( 2 \cosh (\beta
\sum_{\mu\in T_i}u_i^\mu )) \nonumber\\ & &
-\frac{1}{\beta N} \sum_{t=1}^M
 \left[  
E
\log \left(2^{-p}
\sum_{S_1,...,S_p}\e^{-\beta
H_{J^{(t)}}(S_{1},...,S_{_p})}
\prod_{l=1}^p (1+S_l \tanh(\beta g_{i_l^t}^t))
\right)
\right.
\nonumber \\ & &
\left.
-
E\sum_{l=1}^p\left(
\log(\cosh(\beta u_{i_l^t}^t))
+
\log(1+\tanh(\beta g_{i_l^t}^t)\tanh(\beta u_{i_l^t}^t))
\right)
\right]
\label{33}
\end{eqnarray}
from the remainders
\begin{eqnarray}
 R[G,t] & = &
-T\left[
E\left(\log \omega_{-u^t}^{(t-1)}(\e^{-\beta
H_{J^{(t)}}(S_{i^t_1},...,S_{i^t_p})})
\right)
-
E\left( \sum_{l=1}^p \log
( 1+\omega_{-u^t_{i_l^t}}^{(t-1)}(S_{i_l^t})\tanh(\beta u_{i_l^t}^t))
\right)
\right.
\nonumber\\ & &
\left.
-
E
\log \left(2^{-p}
\sum_{S_1,...,S_p}\e^{-\beta
H_{J^{(t)}}(S_{1},...,S_{_p})}
\prod_{l=1}^p (1+S_l \tanh(\beta g_{i_l^t}^t))
\right)
+
E\sum_{l=1}^p\left(\log(1+\tanh(\beta g_{i_l^t}^t)
\tanh(\beta u_{i_l^t}^t))\right)
\right]
\label{34}
\end{eqnarray}
and
\begin{eqnarray}
{\tilde
R}[G,t]=T
\left[
E\left( \log \omega_{-u^t}^{(t-1)}(\prod_{l=1}^p(
(1+S_{i_l^t}\tanh(\beta u_{i_l^t}^t) )
)
\right)
-
E\left(\sum_{l=1}^p \log (
(1+\omega_{-u^t_{i_l^t}}^{(t-1)}(S_{i_l^t})\tanh(\beta u_{i_l^t}^t) 
)
\right)
\right].
\label{35}
\end{eqnarray}
Expressions (\ref{33},\ref{34},\ref{35}) are suggestive on how
to use the interpolating model method to analyze single samples
\cite{FLZ}. Moreover,  we can simplify them observing that
\begin{itemize}
\item
All the $u$ and $g$ fields appearing are statistically independent and
one can therefore assign them arbitrary indexes. 
\item
The marginal of the index probability (\ref{indici}) with respect to
the indexes appearing in the $t$-th clause is the uniform distribution
on $\{1,...,N\}^p$. The indexes appearing in the averages in 
 (\ref{33},\ref{34},\ref{35}) are therefore immaterial. 
\end{itemize}
We finally find: 
\begin{eqnarray}
\beta  F_{var}[G]& = &
 -E\log ( 2 \cosh (\beta
\sum_{l=1}^k u_l )) \nonumber\\ & &
- \alpha 
 \left[  
E
\log \left(2^{-p}
\sum_{S_1,...,S_p}\e^{-\beta
H_{J}(S_{1},...,S_{_p})}
\prod_{l=1}^p (1+S_l \tanh(\beta g_{l}))
\right)
\right.
\nonumber \\ & &
\left.
-
E\sum_{l=1}^p\left(
\log(\cosh(\beta u_{l}))
+
\log(1+\tanh(\beta g_{l})\tanh(\beta u_{l}))
\right)
\right]
\label{33a}
\end{eqnarray}

\begin{eqnarray}
 R[G,t] & = &
-T\left[
E\left(\log \omega_{-u^t}^{(t-1)}(\e^{-\beta
H_{J}(S_{1},...,S_{p})})
\right)
-
p E\left( \log 
(1+\omega_{-u^t_{i_1^t}}^{(t-1)}(S_{i^t_1})\tanh(\beta u) 
)
\right)
\right.
\nonumber\\ & &
\left.
-
E
\log \left( 2^{-p}
\sum_{S_1,...,S_p}\e^{-\beta
H_{J}(S_{1},...,S_{_p})}
\prod_{l=1}^p (1+S_l \tanh(\beta g_{l}))
\right)
+
p E \left(\log(1+\tanh(\beta g)
\tanh(\beta u))\right)
\right]
\label{34a}
\end{eqnarray}
\begin{eqnarray}
{\tilde
R}[G,t]=T
\left[
E\left( \log \omega_{-u^t}^{(t-1)}(\prod_{l=1}^p(
1+S_{i^t_l}\tanh(\beta u_l)))\right)
-p E\left(  \log (
(1+\omega_{-u^t_{i_1^t}}^{(t-1)}(S_{i^t_1})\tanh(\beta u) 
))
\right)
\right]sfto .
\label{35b}
\end{eqnarray}
Direct inspection upon optimizing over the $G$ function shows that
$F_{var}[G]$ coincides with the free energy found with the
replica/cavity method in the replica symmetric approximation.  The
first remainder term $R[G,t]$ is analogous to the one one finds on
Poissonian degree graphs and will be dealt with in section 7.
Conversely, ${\tilde R}$ was absent in the Poissonian case
thanks to the possibility of compensating in average the clause removal
procedure and represents a new difficulty for graphs of arbitrary
connectivity. At a first sight it would seem difficult to say
anything in general about its behaviour. However, we will see in the next
section that, thanks to the  self-averaging property of extensive
quantities, this term becomes vanishingly small in the
thermodynamic limit. This is the only point, besides trivial $O(1/N)$
terms neglection  where the large $N$ limit
enters in our estimates.

\section{Thermodynamic limit}

In order to analyze the cut and paste algorithm let us introduce
$Z(M_1,M_2,M_b)$ as the partition function of the interpolating model
when the number of various kind of clauses are given by $M_1,M_2,M_b$,
and $\omega_{(M_1,M_2,M_b)}(\cdot)$ as the corresponding Gibbs average.
A step of the algorithm, at generic time $t$,  sends $Z(M_1,M_2,M_b)$ to
$Z(M_1+1,M_2+1,M_b-2)$. Let us consider $s$ consecutive steps.
If we denote as $i_1^r,i_2^r$ and $j_1^r,j_2^r$, $r=1,\ldots,s$ the
indexes of the bridge clauses we cut and as $J^1_r$ and $J^2_r$
the corresponding couplings, we can write: 
\begin{eqnarray}
Z(M_1,M_2,M_b)&=& Z(M_1,M_2,M_b-2s)\omega_{(M_1,M_2,M_b-2s)}(\e^{\beta
\sum_r(
  J^1_r S_{i_1^r}S_{i_2^r}+  J^2_r S_{j_1^r}S_{j_2^r})})
\nonumber
\\
Z(M_1+s,M_2+s,M_b-2s)&=& Z(M_1,M_2,M_b-2s)\omega_{(M_1,M_2,M_b-2s)}(\e^{\beta
\sum_r(  J^1_r S_{i_1^r}S_{j_1^r}+  J^2_r S_{i_2^r}S_{j_2^r})}).
\end{eqnarray}
The total average free energy change $\Delta^{(s)}_N$ after the $s$ steps  
is given therefore by: 
\begin{eqnarray}
\beta N \Delta^{(s)}_N &=&\beta N \sum_{\tau=t}^{t+s-1}
\Delta'_N=E\left( \log \omega_{(M_1,M_2,M_b-2s)}(
\prod_r(1+
\tanh(\beta
  J^1_r) S_{i_1^r}S_{i_2^r})(1+\tanh(\beta  J^2_r) S_{j_1^r}S_{j_2^r})) 
\right.
\nonumber\\
&& \left. 
-\log  
\omega_{(M_1,M_2,M_b-2s)}(
\prod_r(1+
\tanh(\beta
  J^1_r) S_{i_1^r}S_{j_1^r})(1+\tanh(\beta  J^2_r) S_{i_2^r}S_{j_2^r}))  
\right).
\label{iiiih!}
\end{eqnarray}
Similarly to  section 4 we add and subtract suitable terms to
$\Delta^{(s)}_N$, to write: 
\begin{eqnarray}
\beta N \Delta^{(s)}_N =&& 2s E\left\{ \log \omega_{(M_1,M_2,M_b-2s)}
(1+
\tanh(\beta
  J^1) S_{i_1}S_{i_2})  \right\}
\nonumber\\
&&  
-sE\left\{\log 
\omega_{(M_1,M_2,M_b-2s)}
(1+
\tanh(\beta
  J^1) S_{i_1}S_{j_1}) 
\right\}-
s E\left\{ \log\omega_{(M_1,M_2,M_b-2s)}(1+\tanh(\beta  J^2) S_{i_2}S_{j_2})
\right\}+\tilde T
\end{eqnarray}
where, for symmetry reasons, we have omitted the index $r$.
Since we have factored the Gibbs averages, the 
remainder $\tilde T$ will contain terms of the type
\begin{eqnarray}
 && E\left[\log  
\omega_{(M_1,M_2,M_b-2s)}(
\prod_r(1+
\tanh(\beta
  J^1_r) S_{i_1^r}S_{j_1^r})(1+\tanh(\beta  J^2_r) S_{i_2^r}S_{j_2^r}))
\right. \\\nonumber
&&\left.-\sum_r\log  
\omega_{(M_1,M_2,M_b-2s)}
(1+\tanh(\beta J^1_r) S_{i_1^r}S_{j_1^r})-\sum_r\log  \omega_{(M_1,M_2,M_b-2s)}
(1+\tanh(\beta  J^2_r) S_{i_2^r}S_{j_2^r}) \right].
\end{eqnarray}
As we will briefly discuss in section 6, the same kind of
self-averaging properties that will ensure the vanishing of $\tilde R$
in the thermodynamic limit, will also guarantee that $\tilde T\to0$.
At this point, as in I, one expands the logarithms in absolutely
convergent Taylor series of the variables $\tanh(\cdots)$. 
Recalling
the fact that the distribution of the $J$ variables is even, one obtains
\begin{eqnarray}
\label{hp}
  \beta N \Delta^{(s)}_N=s
\sum_{n=1}^\infty \frac{\langle\tanh^{2n}\beta J\rangle
}{2n}
E\left[(\omega_{(M_1,M_2,M_b-2s)}(S_{i_1}S_{j_1}))^{2n}-2
(\omega_{(M_1,M_2,M_b-2s)}(S_{i_1}S_{i_2}))^{2n}+
(\omega_{(M_1,M_2,M_b-2s)}(S_{i_2}S_{j_2}))^{2n}
\right].
\end{eqnarray}
Define ${\mathcal I}_1$, ${\mathcal I}_2$  as the 
set of $2s$ indices in the first and second sub-system, respectively,
corresponding to the deleted bridge clauses.
Then it is easy to realize
that, conditionally only on all the non-deleted clauses, the random indices
$i_1,j_1$ and $i_2,j_2$ 
are independently and uniformly distributed on 
${\mathcal I}_1$ and ${\mathcal I}_2$, respectively, 
apart from an error term of order $O(1/s)$. This error term
arises because of the constraints $i_1\ne i_2$, $j_1\ne j_2$, which 
becomes weak for $s$ large, since the number of non-diagonal configurations is 
$O(s^2)$, while that of the diagonal ones is $O(s)$.
 Introducing, in analogy with section 2,
 $\Omega_{(M_1,M_2,M_b-2s)}$ as the replicated version of the measure 
 $\omega_{(M_1,M_2,M_b-2s)}$ and $\tilde q_1^{(2n)}, \tilde q_2^{(2n)}$ as
 \begin{eqnarray}
 &&\tilde q_1^{(2n)}=\frac1{2s} \sum_{i\in {\mathcal I}_1}S^1_{i}
\ldots S^{2n}_{i}\\
 &&\tilde q_2^{(2n)}=\frac1{2s} \sum_{i\in {\mathcal I}_2}S^1_{i}
\ldots S^{2n}_{i},
 \end{eqnarray}
$\Delta^{(s)}_N$ can be rewritten as
 \begin{eqnarray}
 \label{different}
   \beta N \Delta^{(s)}_N=s\sum_{n=1}^\infty \frac{\langle\tanh^{2n}(\beta J)
\rangle}{2n}
 E \,\Omega_{(M_1,M_2,M_b-2s)}\left[(\tilde q_1^{(2n)}-\tilde q_2^{(2n)})^2
\right]+O(1).
\end{eqnarray}
The first term is clearly positive.
As for the term $O(1)$, it becomes negligible in the limit of large $s$,
since the sum in (\ref{inanalogy}) contains $O(N/s)$ terms of this kind and 
has a pre-factor $1/N$.

Notice that this expression is different from the one found in the
procedures used in the Sherrington-Kirkpatrick
\cite{guerraTD} and in the Poissonian
\cite{primolavoro} cases where, in the analogous expansions, 
one finds polynomials in the usual multi-overlaps
$q_1^{(2n)}, q_2^{(2n)}$ defined in section 2, 
depending explicitly on the ratio $N_1/N$.

\section{Consequences of a self-averaging property.}

Formulae (\ref{33a},\ref{34a},\ref{35b}) provide an exact
representation of the free energy. We will see in the next section that 
at least for even $p$, 
the remainder $R[G,t]$ is non negative for all distributions $G$ for
which it makes sense. This is not enough to prove that
$F$ is limited from above by the replica free energy; a control of the 
term ${\tilde R}[G,t]$ is also required. 
As we show below, this control is guaranteed by the self-averaging
property of suitable extensive quantities, which follows from general
thermodynamical convexity arguments, first employed in the context of 
mean field spin glasses by Guerra in \cite{guerraSA}.

As a preliminary fact, notice that, as it is clear
from 
(\ref{FveraRS}), in order to establish the lower bound for the free energy
it is sufficient to show that $\tilde R[G,t]$ vanishes
for $t\le M-\epsilon N$, for arbitrary  $\epsilon>0$ which will be let tend to 
zero in the end.

Let us consider the interpolating model at time $t-1$, and rewrite its 
Hamiltonian (\ref{compoundH}) as
\begin{equation}
\label{calacca}
{\cal H}={\cal H}_0-\sum_{\mu=t}^M \sum_{l=1}^p
u_{i_l^\mu}^\mu S_{i_l^\mu},
\end{equation}
where ${\cal H}_0$ contains the clauses indexed $1,\ldots,t-1$. 
Let us now consider the quantities
\begin{equation}
\phi_k=\frac{1}{N} \sum_{\mu=t}^M\sum_{l=1}^p (u_{i_l^\mu}^\mu)^k S_{i_l^\mu}
\end{equation}
which can be expected to be self-averaging with respect to
the Boltzmann and the quenched averages, for any integer $k$. 
Indeed, one has that {\em generically}
\begin{equation}
\label{fi-fi}
L(k,l)=\lim_{N\to\infty} L_N(k,l)=\lim_{N\to\infty} 
E(\omega(\phi_k \phi_l)-\omega(\phi_k)\omega(\phi_l))=0,
\end{equation}
where what we mean by ``generically'' will be clarified below. For the
moment,
let us explore the consequences of this.
Notice that we can write 
\begin{equation}
L_N(k,l)
=\frac{1}{N^2} \sum_{\mu,\nu=t}^M\sum_{r,s=1}^p
E\left[(u_{i_r^\mu}^\mu)^k (u_{i_{s}^{\nu}}^{\nu})^l 
\frac{\partial}{\partial \beta
  u_{i_r^\mu}^\mu}
\frac{\partial}{\partial \beta u_{i_{s}^{\nu}}^{\nu}} 
\log \omega_{-u_{i_r^\mu}^\mu-u_{i_{s}^{\nu}}^{\nu}} (\e^{\beta
  ( u_{i_r^\mu}^\mu S_{i_r^\mu}+u_{i_{s}^{\nu}}^{\nu} S_{i_{s}^{\nu}}) })\right].
\label{agh}
\end{equation}
Thanks to the average $E$ the sum
is immaterial and we can write that 
 \begin{equation}
\frac{(M-t)^2}{N^2}E\left[(u_{i_r^\mu}^\mu)^k (u_{i_{s}^{\nu}}^{\nu})^l 
\frac{\partial}{\partial \beta
  u_{i_r^\mu}^\mu}
\frac{\partial}{\partial \beta u_{i_{s}^{\nu}}^{\nu}} 
\log \omega_{-u_{i_r^\mu}^\mu-u_{i_{s}^{\nu}}^{\nu}} (\e^{\beta
  ( u_{i_r^\mu}^\mu S_{i_r^\mu}+u_{i_{s}^{\nu}}^{\nu} S_{i_{s}^{\nu}}) 
})\right]\to 0,
\label{id}
\end{equation}
where $\mu\ne \nu$ or $r\ne s$.
Notice that the pre-factor is of order 1.
Next we notice that, conditionally only on the clauses $1,\ldots,t-1$
which have not been
removed, the random variables $i^\mu_r,i_{s}^{\nu} $ are independent and 
identically distributed as
\begin{equation}
\label{indip}
P(i^\mu_1=i)\propto (k_i-k_i(t)),
\end{equation}
where $k_i$ is the degree of the site $i$ in the original system, and 
$k_i(t)$ is its degree at time $t-1$. 
Of course, $k_i-k_i(t)\ge0$ is just the 
number of deleted clauses which involved the $i$'th spin.
In particular, choosing $\mu=\nu=t$, $r=1,s=2$, one can write
 \begin{equation}
E\left[(u_{i_1^t}^t)^k (u_{i_{2}^{t}}^{t})^l 
\frac{\partial}{\partial \beta
  u_{i_1^t}^t}
\frac{\partial}{\partial \beta u_{i_{2}^{t}}^{t}} 
\log \omega_{-u_{i_1^t}^t-u_{i_{2}^{t}}^{t}} (\e^{\beta
  ( u_{i_1^t}^t S_{i_1^t}+u_{i_{2}^{t}}^{t} S_{i_2^t}) })\right]\to 0.
\label{idbis}
\end{equation}
Observing that this identity
 has to be valid for all $k$ and $l$, we find that
with probability 1 with respect to the distribution of $u^t_{i^t_1}$ and $u^t_{i^t_2}$,  
 \begin{equation}
 \frac{\partial}{\partial \beta
  u_{i_1^t}^t}
\frac{\partial}{\partial \beta u_{i_{2}^{t}}^{t}} 
E_{-u_{i_1^t}^t-u_{i_2^t}^t}\left(\log \omega_{-u_{i_1^t}^t-u_{i_2^t}^t} 
(\e^{\beta  (u_{i_1^t}^t S_{i_1^t}+u_{i_2^t}^t S_{i_2^t}) }\right)\to 0.
\label{id0}
\end{equation}
This implies that 
 \begin{eqnarray}\nonumber
  E_{-u_{i_1^t}^t-u^t_{i^t_2}}\left(\log \omega_{-u_{i_1^t}^t-u^t_{i^t_2}} (\e^{\beta
  (u_{i_1^t}^t S_{i_1^t}+u^t_{i^t_2} S_{i_2^t}) })\right)&\approx&
  E_{-u_{i_1^t}^t-u^t_{i^t_2}}\left(\log \omega_{-u_{i_1^t}^t-u^t_{i^t_2}} (\e^{\beta
  u_{i_1^t}^t S_{i_1^t} })\right)
\\&&+
  E_{-u_{i_1^t}^t-u^t_{i^t_2}}\left(\log \omega_{-u_{i_1^t}^t-u^t_{i^t_2}} (\e^{\beta
  u^t_{i^t_2} S_{i_2^t} })\right)
\label{id1}
\end{eqnarray}
and that
 \begin{eqnarray}
  E_{-u_{i_1^t}^t -u^t_{i^t_2}}\left(\log \omega_{-u_{i_1^t}^t} (\e^{\beta
  u_{i_1^t}^t S_{i_1^t} })\right)&=&
  E_{-u_{i_1^t}^t-u^t_{i^t_2}}\left(\log \omega_{-u_{i_1^t}^t-u^t_{i^t_2}} (\e^{\beta
  u_{i_1^t}^t S_{i_1^t} +u^t_{i^t_2} S_{i_2^t}})\right)
-
  E_{-u_{i_1^t}^t-u^t_{i^t_2}}\left(\log \omega_{-u_{i_1^t}^t-u^t_{i^t_2}} (\e^{\beta
  u^t_{i^t_2} S_{i_2^t} })\right)
\nonumber\\
&\approx & E_{-u_{i_1^t}^t-u^t_{i^t_2}}\left(\log \omega_{-u_{i_1^t}^t-u^t_{i^t_2}} (\e^{\beta
  u_{i_1^t}^t S_{i_1^t} })\right)
\label{id2}
\end{eqnarray}
where the equivalence sign $\approx$ means in the previous formulae
and in the rest of the section that both quantities on the two sides of
an equation tend to the same value in the thermodynamic limit. It is
easy to realize that these identities can be generalized to ones 
involving arbitrary number $j$ of fields: 
\begin{equation}
  E_{-u^t_{i^t_1},...,-u^t_{u^t_j}}\left(\log \omega_{-u^t_{i^t_1}
,...,-u^t_{i^t_j}}
  (\e^{\beta \sum_{r=1}^j u^t_{i^t_r} S_{i^t_r} })\right)\approx 
\sum_{r=1}^j
  E_{-u^t_{i^t_1},...,-u^t_{u^t_j}}
\left(\log \omega_{-u^t_{i^t_r}}
  (\e^{\beta u_{i^t_{i^t_r}} S_{i^t_r}})\right).
\label{id3}
\end{equation}
If we now take $j=p$ and recall (\ref{35}),  we conclude that self-averaging of
the $\phi_k$ implies that ${\tilde R}[G,t]$ tends to zero in the
thermodynamic limit.

Similarly, to prove that the remainder $\tilde T$ of section 5 vanishes,
one has to consider a Hamiltonian of the form
 $${\cal H}={\cal H}_0-\sum_{\mu=1}^M w_\mu S_{i_1^\mu}S_{i_2^\mu}
$$
and to exploit self-averaging of the quantities 
$$
\psi_k=\frac1N\sum_{\mu}w^k_\mu S_{i_1^\mu}S_{i_2^\mu},
$$
%where $w_\mu$ are suitably defined fields. We do not give details here.
Repeating the steps which led from (\ref{fi-fi}) to (\ref{id3}), 
one finally finds that $\tilde T$ vanishes for $N\to\infty$.

We are now left with the task of showing that relations like
(\ref{fi-fi}) generically hold. The strategy we use has been already
employed in \cite{guerraSA}, and more recently in \cite{T}.  The idea
is as follows: while one is not able in general to prove that
Eqs. (\ref{fi-fi}) hold for the original model (\ref{calacca}), one
can perturb it by adding a term $N \sum_{k=1}^\infty \lambda_k \phi_k$
to the Hamiltonian ${\cal H}$, $\lambda_k$ being real numbers
decreasing sufficiently fast with $k$, so that the infinite volume
free energy remains bounded.\footnote{Results similar to those described 
below could be
obtained considering perturbations of the kind $N^{1-\alpha}
\sum_{k=1}^\infty \lambda_k \phi_k$ with $0<\alpha<1/2$, which even
for finite $\lambda$'s modify the free energy or equations (\ref{agh})
only by $O(N^{-\alpha})$ terms.} In the end, we will be interested in
the case where all the $\lambda$'s vanish.  Then, as in
\cite{guerraSA}, convexity of the free energy with respect to
parameters $\lambda_k$ implies that, for almost every choice for their
values, the second derivative $ \partial^2_{\lambda_k} F_N $ is
finite, also in the thermodynamic limit. Since this derivative can be
written as
$$
\partial^2_{\lambda_k} F_N=N E\left(\omega_\lambda(\phi^2_k)-
\omega_\lambda(\phi_k)^2\right),
$$
it follows that equations (\ref{fi-fi}) hold, for almost every choice of 
the $\lambda$'s, if the Gibbs averages are understood to 
correspond to the modified model. 
In order to conclude the argument and show that $\tilde R\to0$, 
we have to remark two important points. First, Eqs. (\ref{agh}) will be 
modified by a harmless remainder of order $O(\lambda)$, which will disappear
in the end 
since  we can take the $\lambda_k$ arbitrarily small. Next, the remainder
term $R$ of section 4 remains positive, so that Eq. (\ref{FveraRS}) gives
\begin{equation}
\label{bounde}
  F_N\ge F_{var}[G]+O(1/N)+O(\lambda),
\end{equation}
where both the replica-symmetric and the true free energy are those
corresponding to the modified system. Finally, one employs the fact that
the free energy is always a continuous function, as a consequence of
its convexity, to deduce that (\ref{bounde}) holds also for the original 
system with all $\lambda$'s set to $0$.
It is important to emphasize that we have not used anywhere assumptions 
of continuity of Gibbs averages with respect to the parameters $\lambda$, 
which we cannot prove in general,
but just simple positivity properties and the continuity of the free energy,
which holds as a general fact.

\subsubsection{Generalization to multi-overlaps of the replica
  equivalence identities} 

It is interesting to investigate the consequences of Eq. (\ref{id}) on
the distribution of multi-overlaps. To that scope, one can perform an
expansion in powers of $\tanh(\beta v_i)$ and $\tanh(\beta v_j)$ in
(\ref{id}).  Omitting tedious but conceptually simple calculations one
finds that in the thermodynamic limit, for all $r$ and $s$, 
\begin{equation}
\sum_{l=0}^{min[2r,2s]} \frac{(-1)^{(l+1)}
(2r+2s-l-1)!}{(2r-l)!(2s-l)!l!}  \langle (q^{(2r)}\cdot
q^{(2s)})_{l} \rangle = 0 \; ,
\label{overlaps}
\end{equation} 
where the $q^{(2r)}$ and $q^{(2s)}$ are multi-overlaps involving
respectively $2r$ and $2s$ replicas, and with the notation
$(q^{(2r)}\cdot q^{(2s)})_{l}$ we mean that among the two groups of
replica $l$ are in common. Notice that the relations (\ref{overlaps})
generalize to the case of multi-overlaps the Ghirlanda-Guerra identity
 \cite{GG} 
\begin{equation}
-\frac{3}{2} \langle q_{12}q_{34}\rangle +2  \langle q_{12}q_{23}\rangle
-\frac{1}{2} \langle q_{12}^2\rangle
=0
\end{equation} 
and reduce to it for $r=s=1$. 
More general relations can be obtained
from (\ref{id3}) or, in analogy with \cite{GG}, considering
considering self-averaging properties of multi-spin perturbations in
the Hamiltonian. While we will not pursue this route in this paper in
full generality, it is clear that $p$-spin perturbations to the
Hamiltonian will give rise to equations similar to (\ref{overlaps})
involving multi-overlaps raised to the  $p$-th power. 

We notice that -as it happens in the Ghirlanda-Guerra case- 
within the replica method, identities (\ref{overlaps}) can be derived
from the requirement of ``replica equivalence''. 
Within replica method one introduces
multi-overlap order parameters $Q_{a_1,...,a_r}$, and
self-consistently finds that the multi-overlap averages are given by
\begin{equation}
\langle q^{(r)} \rangle 
= \lim_{n \to 0} \frac{1}{n(n-1)...(n-r+1)} \sum_{a_1,...,a_r}^{',1,n} Q_{a_1,...,a_r},
\label{rep}
\end{equation} 
where all the indexes in $\sum^{'}$ are different. Similarly, the
averages $\langle (q^{(r)}\cdot
q^{(s)})_{l} \rangle$ are given by replica sums of
$Q_{a_1,...,a_r}Q_{b_1,...,b_s}$ with $l$ $a$-indexes coinciding with
$l$ $b$-indexes, normalized to the number of terms in the sum. For
instance,  $\langle (q^{(2)}\cdot
q^{(2)})_{1} \rangle=
\frac{1}{n(n-1)(n-2)}\sum_{a,b,c}' Q_{ab}Q_{bc}$. 
 Replica equivalence states that for all values of $a_1$ the sum 
\begin{equation}
\sum_{a_2,...,a_r}^{1,n} Q_{a_1,...,a_r}. 
\end{equation} 
takes the same value. As a consequence, for $n\to 0$
\begin{equation}
\frac{1}{n}\sum_{a_1,...,a_r}^{',1,n} Q_{a_1,...,a_r}\sum_{b_1,...,b_s}^{',1,n} Q_{b_1,...,b_s}
= O(n)\to 0. 
\end{equation}
Singling out in the sum terms with coinciding indexes in the overlaps
and making use of (\ref{rep}), after some algebra one finds
Eq. (\ref{overlaps}). Eq. (\ref{id}) can be seen as the generating
function of these identities. 

\section{The remainder $R[G,t]$.}

While the control of the remainder $\tilde R$ was obtained, in the
previous
section, without making reference to a  specific form for the
Hamiltonian
(\ref{general}), in the present section we restrict ourselves 
for simplicity to the $p$-spin case, where
$H_J(S_{i_1},...,S_{i_p}) = J\,
S_{i_1}\cdot ...\cdot S_{i_p}$.  
Substituting in Eq. (\ref{FveraRS}), 
using the independence among the external fields and the relation 
(\ref{qdiu}) one finds for the first remainder term:
\begin{eqnarray}
R^{p-spin}[G,t]=
 & & -\frac{1}{\beta}
 \left[ E
  \left\langle 
          \log(1+\tanh(\beta J)\omega^{(t-1)}_{-u^t}
            (S_{i^t_1}...S_{i_p^t}))
  \right\rangle_J 
-p E 
  \left\langle 
\log(1+\tanh (\beta
  J)\prod_{l=1}^{p-1}\tanh(\beta g_l)
  \omega^{(t-1)}_{-u^t_{i_1^t}} (S_{i_1^t}))
\right\rangle_J  + 
\nonumber \right. \\
  & & \left. (p-1) E
  \left\langle 
 \left(  \log(1+\tanh (\beta
  J)\prod_{l=1}^p \tanh (\beta g_l))
  \right) 
 \right\rangle_J \right].
\end{eqnarray}
Through this expression we can establish that the remainder is
positive for even $p$.  

Now, we expand the logarithm of the three terms into
(absolutely convergent) series of $\tanh(\beta J)$, and notice that
thanks to the parity of the $J$ and the $g$ distributions, they will
just involve negative terms. We can then take the expected value of
each term and write
\begin{eqnarray}
R^{p-spin}[G,t]= \frac{1}{\beta}  \sum_{n=1}^{\infty} 
 \frac{\langle \tanh^{2 n} \beta J \rangle_{J}}{2n}
E\left[(\omega^{(t-1)}_{-u^t}(S_{i^t_1}\ldots S_{i^t_p}))^{2n}-p
\langle \tanh^{2n}\beta g\rangle_g^{p-1}(\omega^{(t-1)}_{-u^t_{i_1^t}}(S_{i^t_1}))^{2n}
+(p-1)\langle  \tanh ^{2 n} \beta g \rangle_g^p
\right].
\end{eqnarray}
In analogy with 
section 2, we introduce $\Omega^{(t-1)}$ as the replicated version of 
the measure $\omega^{(t-1)}$.
Next we notice that, as in section 6, conditionally only on the clauses $1,\ldots,t-1$
which have not been
removed, the random variable $i^t_l$ is distributed as
\begin{equation}
P(i^t_l=i)\propto (k_i-k_i(t)).
\end{equation}
Then, defining
\begin{equation}
\hat\psi_i^a=\e^{-\beta u^1_i S^a_i}\frac{Z(t-1)}{Z_{-u^1_i}(t-1)}
\end{equation}
and
\begin{equation}
r^{(2n)}=\frac{\sum_i(k_i-k_i(t))S^{1}_i\hat\psi^1_i\ldots S^{2n}_i
\hat\psi^{2n}_i}{\sum_i(k_i-k_i(t))}=
\frac{\sum_i(k_i-k_i(t))S^{1}_i\hat\psi^1_i\ldots S^{2n}_i\hat\psi^{2n}_i}
{p(M-t+1)},
\end{equation}
it is easy to realize that
\begin{equation}
R^{p-spin}[G,t]=  \frac{1}{\beta}  \sum_{n=1}^{\infty} 
 \frac{\langle \tanh^{2 n} \beta J \rangle_{J}}{2n} E\,\Omega^{(t-1)}
\left[
(r^{(2n)})^p - p\,r^{(2n)}  \langle \tanh ^{2 n} \beta g \rangle_g^{p-1} + 
(p-1) \langle  \tanh ^{2 n} \beta g \rangle_g^p \right]+O(1/N).
\label{sser}
\end{equation}
The observable $\hat \psi_i$ can be seen as the operator which annihilates 
one of the  external fields $u_i$ from  site $i$. Indeed, for any observable
$A$ one has
\begin{eqnarray}
\omega (A\hat \psi_i)=\omega_{-u^1_i}(A).
\end{eqnarray}
The harmless error term $O(1/N)$ arises because, in order to reconstruct $(r^{(2n)})^p$, 
we added ``diagonal'' terms
where at least two indexes $i^t_{l}$
and $i^t_{l'}$ are equal. Since $M-t>\epsilon N$ with $\epsilon>0$, as in
section 6, these 
terms give altogether a vanishing contribution in the infinite volume limit.
It is interesting to notice that, as found in I, for models with
Poissonian connectivity degree the expansion of $R[G,t]$ can be expressed
in terms of the usual multi-overlap $q^{(2n)}$ defined in section 2.

In the case of the $K$-SAT, using definition (\ref{SAT}) for the clause $H_J$,
we find  relation:
\begin{equation}
u_{\bf J}(g_1,...,g_{p-1}) \equiv u_J(\{J_l\},\{g_l\}) =
\frac{J}{\beta} \tanh^{-1} \left[  \frac{\frac{\xi}{2}
\prod_{l=1}^{p-1} \left( \frac{1 + J_l \tanh( \beta g_l )}{2}
\right)}{1 +\frac{\xi}{2} \prod_{l=1}^{p-1} \left( \frac{1 + J_l
\tanh( \beta g_l )}{2}  \right) } \right] \; ,
\label{U-KSAT}
\end{equation}
where $\xi \equiv e^{-\beta} -1 < 0$. 
Therefore, one has
\begin{eqnarray}
R^{K-SAT}[G,t] &=& -\frac{1}{\beta} E \left[  \left\langle \log
\left(  1+(e^{-\beta}-1)\omega(\prod_{l=1}^p \frac{1+J_{l} S_{i^t_l}}{2})
\right) \right\rangle_{\{J_l\}}  \right. \nonumber \\ & & -p
\left\langle \log \left(  1+ \xi \omega\left(\frac{1+J S_{i^t_1}}{2}
\prod_{l=1}^{p-1}\frac{1+J_{l} \tanh (\beta g_l)}{2} \right)  \right)
\right\rangle_{\{g_l\}, J, \{J_l\}} + \nonumber \\ & & \left. (p-1)
\left\langle \log\left( 1+ \xi \prod_{l=1}^{p}\frac{1+J_{l} \tanh
(\beta g_l)}{2}\right) \right\rangle_{\{g_l\}, \{J_l\}} \right] \; .
\label{RESTOKSAT}   
\end{eqnarray}
Expanding in series the logarithms, exploiting
the symmetry of the probability distribution functions and taking
the expectation of each term of the absolutely convergent series we
obtain:
\begin{equation}
R^{K-SAT}[G,t] = \frac{1}{\beta} \sum_{n\ge 1}
\frac{(-\xi^*)^{n}}{n} \Omega \left[ (1+R_n)^p -
p(1+R_n)\langle(1+J \tanh (\beta g))^n\rangle_{J,g}^{p-1} +\\(p-1)
\langle(1+J \tanh (\beta g))^n\rangle_{J,g}^{p} \right]
\label{RESTO2KSAT}
\end{equation}
where we have defined $\xi^* \equiv \xi/(2^p) < 0$ and $R_n \equiv
\sum_{l=1}^n \langle J^l \rangle_J \sum_{a_1<...<a_l}^{1,n}
r^{a_1...a_l}$.  

We notice that both in the $p$-spin and in the K-SAT problem 
the remainder can be
written as a series of the type
\begin{equation}
R[G,t] =\sum_{n=1}^{\infty} C_n E\,\Omega [
f_p(X_{n},Y_{n}) ] \; ,
\label{general-rem}
\end{equation}
where
\begin{equation}
f_p(x,y) =  x^p - p x y^{p-1}  + (p-1)
y^p \; ,
\label{general-rem2}
\end{equation}
$C_n \ge 0$ for any temperature, $X_n$'s are suitable combinations of
overlaps and $Y_n$'s averages of $g$-fields hyperbolic tangents
moments that correctly calculate the overlaps in the corresponding
replica symmetric approximation.
More specifically, for the $p$-spin 
\begin{eqnarray}
C_n& =& \frac{\langle \tanh^{2n}(\beta J)\rangle}{2\beta n}
\nonumber\\
X_n&=&r^{(2n)}
\nonumber\\
Y_n&=&\langle \tanh^{2n}(\beta g)\rangle
\end{eqnarray}
while for the K-SAT
\begin{eqnarray}
C_n& =& \frac{ (-\xi^*)^n}{\beta n}
\nonumber\\
X_n&=&1+R_n
\nonumber\\
Y_n&=&\langle (1+J\tanh(\beta g))^n\rangle. 
\end{eqnarray}
As noticed in I,  for even $p$ the function $f_p(x,y)$
is positive for all real $x$ and $y$ thus ensuring the positivity of
the remainder. 
For odd $p$ this is not the case as
$f_p(x,y)$ at fixed $y$ becomes negative for $x$ negative and large
enough. Although physically one expects that for odd $p$ negative
values of $X_n$ appear with probability exponentially small in $N$, we
have not been able to prove this property in full generality. 

The odd $p$ case is however interesting. In particular, one would like
to be able to control the remainder in the case of the 3-SAT problem,
recently solved within 1RSB scheme \cite{PRE}. When explicit solutions
of the replica equations exist, one can try to plug them into the
expression of the remainder, to check positivity. Results in this directions,
in the RS case at zero temperature for K-SAT and $p$-spin models, will
be reported in Ref. \cite{inpreparation}.

\section{Summary and conclusions}

In this paper we have shown that the free energy of diluted
spin glass models with arbitrary random connectivity can be written as the sum of a term identical to
the ones got in the cavity/replica plus an error term. The expression
has been obtained through the introduction of an auxiliary model
interpolating between the original model and a pure paramagnet. The
interpolation can be though of as a discrete time dynamical process 
in which the terms
of the Hamiltonian are progressively removed, while the removal effect
is compensated by the introduction of some external fields. The
procedure generalizes the previous work \cite{primolavoro} 
on Poissonian graphs, where the
compensation could be performed in average: at each step one adds
there a random number of fields on random sites. In the present case,
on the other hand, a detailed
compensation where one puts a field on each site involved in the
erased clause is necessary. As a consequence, a new term in the
remainder appears. We have shown that, thanks to self-averaging of suitable
extensive quantities, this new term gives a vanishing
contribution in the thermodynamic limit. 
The rest of the remainder is manifestly positive for even
$p$.

It is also possible to show, as it was done in \cite{guerraVB} 
for models with Poissonian random connectivity, that the free energy 
and the ground state energy are self-averaging quantities, and one can
obtain upper bounds, exponentially small in the system size, for the 
probability of large fluctuations.

\vspace{.5cm}
{\bf Acknowledgements}

We are very grateful to Andrea Montanari for pointing out a weak point 
in the 
first version of this work.

\end{document}